\documentclass[twocolumn,showpacs,preprintnumbers,amsmath,amssymb,aps,prl]{revtex4}
\usepackage{graphicx}
\begin{document}

\title{Transverse Commensurability Effect For Vortices in Periodic Pinning Arrays}   
\author{C. Reichhardt and    
 C. J. Olson Reichhardt} 
\affiliation{
 Theoretical Division,
Los Alamos National Laboratory, Los Alamos, New Mexico 87545 } 

\date{\today}
\begin{abstract}
Using computer simulations, we demonstrate a new type of 
commensurability 
that occurs for vortices moving longitudinally through periodic pinning arrays
in the presence of an additional transverse driving force.
As a function of vortex density, there is
a series of broad maxima in the transverse critical depinning force 
that do not fall at the matching fields 
where the number of vortices equals an integer multiple of the 
number of pinning sites. The commensurability effects are 
associated with 
dynamical states 
in which evenly spaced structures 
consisting of one or more moving rows of vortices form between rows
of pinning sites.
Remarkably, the critical transverse depinning force 
can be more than an order of magnitude larger than the
longitudinal depinning force. 
\end{abstract}
\pacs{74.25.Qt}
\maketitle

\vskip2pc
Matching effects for vortices in 
periodic pinning arrays have been studied extensively for 
different types of pinning lattice geometries 
\cite{Baert,Harada,Reichhardt,Peeters,Schuller,Rosseel}.
As a function of 
magnetic field, 
the critical current passes through a series of peaks 
generated by commensurability effects that occur
when the number of vortices equals an integer 
multiple of the number of pinning sites and the vortex
ground state is an ordered crystalline structure
\cite{Harada,Reichhardt,Peeters}.
When each pinning site can capture only one vortex, 
the excess vortices at fields above the first matching field
are located in interstitial sites and depin 
first under an external drive 
\cite{Harada,Reichhardt,Olson,Peeters,Rosseel}.    
Once the interstitial
vortices are moving under a longitudinal drive, it is possible to
apply an additional transverse drive in the direction perpendicular to
the vortex motion.
In this case, although the vortices are mobile in the longitudinal 
direction, they can remain pinned in the transverse direction and there can
be a finite transverse critical depinning threshold.

The possibility of a transverse depinning threshold for moving 
vortices was initially predicted for 
systems with random pinning when the moving vortices
form well defined channels \cite{Giamarchi}, and transverse
depinning thresholds in randomly pinned systems have been observed
in numerical simulations \cite{Tran} and
experiments \cite{Hilke,Hilke2}. 
For vortices moving in the presence of a periodic 
pinning array, a finite transverse 
depinning threshold 
has been measured at high drives when all of the vortices are moving and
has a value that
depends on the angle between the longitudinal driving direction and
a symmetry axis of the pinning lattice.
Here, the most prominent transverse depinning thresholds and 
dynamical locking effects occur for driving along the
principal axes of the pinning lattice \cite{Nori,Cariero}.
This type of effect has been 
experimentally observed for colloidal particles moving over  
periodic substrates \cite{Korda}. 
When only one vortex can be captured by each pinning site, 
motion of the vortices at low drives
occurs as a flow of interstitial vortices between
vortices that remain trapped at the pinning sites 
\cite{Harada,Reichhardt,Rosseel}.
In this case, it is not known whether
a transverse depinning threshold exists, 
and in general it is not known how the transverse depinning threshold 
varies with magnetic field.   

It might be expected that 
the transverse depinning threshold would simply exhibit peaks at the
same magnetic fields where peaks in the longitudinal depinning threshold
appear.
In this work, we demonstrate that although there are enhancements of the
transverse depinning threshold at certain fields, these fields are
{\it not related} to fields which produce peaks in the longitudinal
depinning threshold, but are instead associated with {\it dynamical}
matching conditions.
The distinct dynamical matching effects appear because the moving vortices
assume a different structure than the static vortex ground state.
Dynamical commensurability effects occur when
an integer number of moving interstitial vortex rows 
form between adjacent rows of pinning sites.
The dynamical matching effects are much broader than the static matching
effects and have maxima that encompass several static matching fields.
An oscillatory critical current appears
for the dynamical transverse commensurability effect. 
This is similar to the critical current oscillations 
seen for vortices depinning in artificial channels \cite{Kes,Anders}
or in layered or strip geometries \cite{Griessen,Souza},
although in the channel, layer, or strip systems, the commensurations arise
due to matching effects of the vortex ground state rather than the dynamical
matching effects observed in the present work. 
Remarkably, we find that the transverse depinning threshold can be 
up to an order of magnitude larger than the longitudinal depinning threshold.
   
We numerically simulate 
a two-dimensional system 
with periodic boundary conditions in the
$x$ and $y$ directions containing 
$N_{v}$ vortices and $N_{p}$ pinning sites
following a procedure similar to that 
used in previous simulations for vortices in 
periodic pinning arrays. The   
number of vortices is proportional to the applied magnetic field 
${\bf B}=B{\bf {\hat z}}$, which is normal to our simulation plane.
The repulsive vortex-vortex interaction force
is given by
${\bf F}^{vv}_{i} = \sum^{N_{v}}_{i\neq j}f_{0}K_{1}(R_{ij}/\lambda){\hat {\bf R}}_{ij}$,  
where $K_{1}$ is a modified Bessel function, 
$R_{ij}=|{\bf R}_i-{\bf R}_j|$ 
is the distance between vortex $i$ and $j$ located at ${\bf R}_i$ and 
${\bf R}_j$, ${\hat {\bf R}}_{ij}=({\bf R}_i-{\bf R}_j)/R_{ij}$, 
$f_{0} = \phi_{0}^2/(2\pi\mu_{0}\lambda^3)$,
$\phi = h/2e$ is the elementary flux quantum, 
and $\lambda$ is the London penetration depth. 
The pinning sites are placed in a triangular lattice, and the field 
at which the number of pinning sites equals the number of vortices,
$N_p=N_v$, is defined as the matching field $B_{\phi}$.     
The individual pinning sites are  modeled as 
parabolic traps
of radius $r_{p} = 0.35\lambda$ and strength $F^{p}=1.25$,
with 
${\bf F}^p_i=\sum_k^{N_p}F^pf_0(R_{ik}/r_p)\Theta(r_p-R_{ik}){\hat {\bf R}}_{ik}$,
where $\Theta$ is the Heaviside step function, $R_{ik}=|{\bf R}_i-{\bf R}^p_k|$,
${\hat {\bf R}}_{ik}=({\bf R}_i-{\bf R}^p_k)/R_{ij}$, and ${\bf R}_k^p$ is
the location of pin $k$.
The overdamped equation of motion for a single vortex $i$ is
\begin{equation} 
\eta\frac{d {\bf R}_{i}}{dt} = {\bf F}^{vv}_i + {\bf F}^{p}_i + {\bf F}^{ext},  
\end{equation} 
where $\eta=1$ is the viscous damping term. 
${\bf F}^{ext}$ represents the net force from
an applied current and is given by 
${\bf F}^{ext}=F^{L}_Df_0{\bf {\hat x}} + F^{Tr}_Df_0{\bf {\hat y}}$,
where the longitudinal drive $F^{L}_D$ is applied in the $x$ direction and 
the transverse drive $F^{Tr}_D$ is applied in the $y$ direction.
The initial vortex positions are 
obtained by simulated annealing.
The drive is first applied in the longitudinal direction 
in increments of $\Delta F^{L}_D=0.0015$, with 15000 simulation time steps
spent at each current increment.
Once the longitudinal drive reaches the desired value, 
it is held fixed while the transverse drive is increased from zero with the
same current increment protocol.
We find that our increment rate is sufficiently slow to avoid any transient
effects.
The longitudinal and transverse critical depinning thresholds,
$F^{L}_c$ and $F^{Tr}_c$, are obtained by measuring the vortex velocity  
$\langle V_{\alpha}\rangle =N_v^{-1}\sum_i^{N_v} {\bf v}_i \cdot {\hat {\bf \alpha}}$, with
$\alpha=x,y$, and identifying the driving force at which
$\langle V_{\alpha}\rangle >0.001$.

\begin{figure}
\includegraphics[width=3.5in]{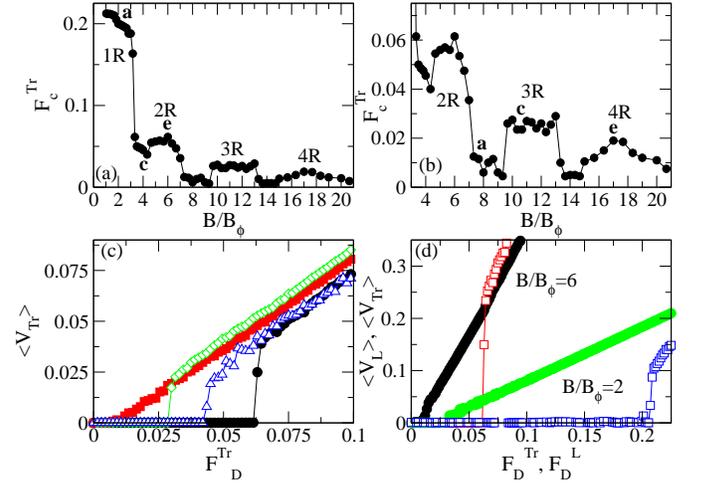}
\caption{
(a) The transverse critical depinning force 
$F^{Tr}_{c}$ vs $B/B_{\phi}$ for a system with 
$B_\phi=0.052\phi_0/\lambda^2$  
and fixed
longitudinal drive 
$F^{L}_{D} = 0.6$. 
The points {\bf a}, {\bf c}, and {\bf e} refer to the fields illustrated
in Fig. 2.
The maxima are labeled according to the number of moving vortex rows between
adjacent pinning rows: 1R, one row; 2R, two rows; 3R, three rows; and 4R,
four rows.
(b) A blow up of panel (a)
for $B/B_{\phi} > 2.0$. 
The points {\bf a}, {\bf c}, and {\bf e} refer to the fields illustrated
in Fig.~3.
(c) The transverse velocity $\langle V_{Tr}\rangle$ vs 
transverse force $F^{Tr}_{D}$ for 
$B/B_{\phi} = 4.33$ 
(open triangles), 
$6.0$ 
(filled circles), 
$8.0$ 
(filled squares),  
and $10$ 
(open diamonds). 
(d) The scaled longitudinal velocity
$\langle V_{L}\rangle (N_v/N_p)$ (filled circles) versus 
longitudinal drive $F_{D}^L$ and 
transverse velocity $\langle V_{Tr}\rangle (N_v/N_p)$ (open squares) 
vs transverse drive $F_{D}^{Tr}$ 
for $B/B_{\phi} = 6.0$ (left curves) and $2.0$ (right curves). 
}
\end{figure}

We first study a system with a low pinning density to 
ensure that a portion of the
vortices are located in the interstitial sites. 
The existence of a clearly defined depinning threshold 
which varies nonmonotonically with field is illustrated
in Fig.~1(c), where we show 
$\langle V_{Tr}\rangle$ versus $F_{D}^{Tr}$ for 
$B/B_{\phi} = 4.33$, $6.0$, $8.0$, and $10.0$ in 
a system with 
$B_{\phi} = 0.052\phi_0/\lambda^{2}$ 
and fixed $F^{L}_D=0.6$. 
From a series of simulations, 
we obtain the variation in $F_c^{Tr}$ versus $B/B_\phi$
plotted in Fig.~1(a).
Four well defined 
maxima in $F^{Tr}_{c}$ appear that are centered near
$B/B_{\phi} = 2.0$, 6.0, 12.0, and 17.0. Figure 1(b)
shows a blowup of the region $B/B_{\phi} > 2.0$, where the oscillation
in $F^{Tr}_c$ can be seen more clearly.
This oscillation is distinct from the matching
effects observed for longitudinal depinning 
\cite{Baert,Reichhardt,Peeters,Schuller}, where well defined peaks occur at  
integer matching fields. 
The maxima in Fig.~1(a) are much broader than in the longitudinal
depinning case and each encompass
three or more
matching fields.  Similarly, the minima in $F_c^{Tr}$ also each spread
over several values of $B/B_{\phi}$.

For all fields $B/B_{\phi} > 1.0$, we find that 
$F_{c}^{Tr}$ is significantly larger
than the longitudinal critical force $F^{L}_{c}$, 
as shown in Fig.~1(d) where we plot 
$\langle V_{L}\rangle(N_v/N_p)$ versus 
$F^{L}_{D}$ and
$\langle V_{Tr}\rangle(N_v/N_p)$ vs $F^{Tr}_{D}$. 
Here the velocities have been scaled by $N_p$ rather than $N_v$ 
for presentation purposes.
For $B_{\phi} = 2.0$, 
the transverse depinning threshold 
$F^{Tr}_c$ is
about six times higher than the longitudinal depinning threshold
$F^{L}_c$. 
Both depinning thresholds are lower
for $B/B_{\phi} = 6.0$;
however, $F^{Tr}_c$ is again much higher than $F^{L}_c$.

\begin{figure}
\includegraphics[width=3.5in]{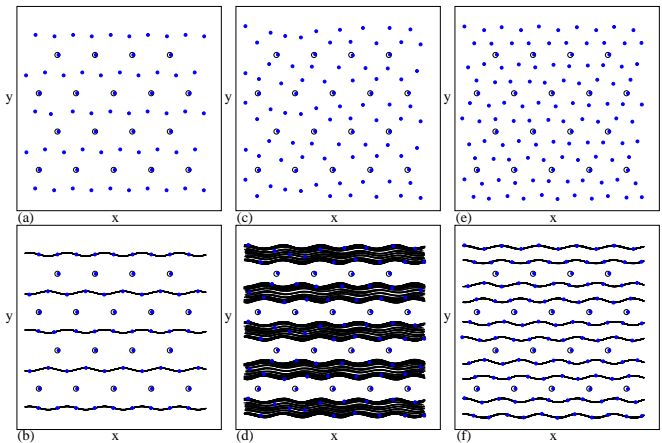}
\caption{
The vortex positions (black dots), pinning site locations (open circles), and 
vortex trajectories (black lines) for the system in Fig.~1(a). 
(a), (b) $B/B_{\phi} =  2.67$, marked {\bf a} in Fig.~1(a). 
(c), (d) $B/B_{\phi} = 4.0$, marked {\bf c} in Fig.~1(a).
(e), (f) $B/B_{\phi} = 6.0$, marked {\bf e} in Fig.~1(a). 
}
\end{figure}

In Fig.~2(a) we show the  
vortex and pinning site positions for 
point {\bf a} in Fig.~1(a) at  $B/B_{\phi} = 2.67$ and in 
Fig.~2(b) we illustrate the vortex trajectories 
for $F_D^{Tr} \lesssim F_c^{Tr}$, just below the transverse depinning transition.
There is a single row of moving interstitial vortices between neighboring rows
of pinning sites
and the vortex lattice is anisotropic, with higher vortex density in the
interstitial rows than in the pinned rows.
The same vortex structure appears for $1.0 < B/B_{\phi} < 2.9$, 
corresponding to the maximum in $F_c^{Tr}$ marked 1R in Fig.~1(a). 
In Fig.~2(c), we plot the vortex positions for 
$B/B_\phi=4.0$ at a minimum of $F^{Tr}_c$ found at 
the point marked {\bf c} in Fig.~1(a).
The interstitial rows are no longer uniform and consist of an interlacing of
double rows with single rows.
Figure 2(d) shows that the vortex trajectories 
at this field are more disordered.
The vortex positions and trajectories at point {\bf e} 
in the region marked 2R in Fig.~1(a) for
$B/B_{\phi} = 6.0$ appear in Fig.~2(e,f).
Here there are two well defined rows of moving
vortices between adjacent pinning site rows.

\begin{figure}
\includegraphics[width=3.5in]{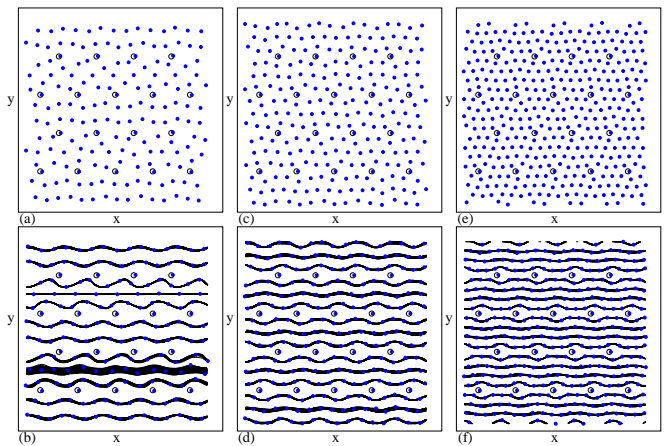}
\caption{ 
The vortex positions (black dots), pinning site 
locations (open circles), and vortex
trajectories (black lines) for the system in Fig.~1(a). 
(a), (b) $B/B_{\phi} = 8.0$, marked {\bf a} in Fig.~1(b).
(c), (d) $B/B_{\phi} = 10.67$, marked {\bf c} in Fig.~1(b). 
(e), (f) $B/B_{\phi} = 17$, marked {\bf e} in Fig.~1(b).
}
\end{figure}

We find that maxima in $F^{Tr}_c$ occur 
whenever there is an integer number of moving 
rows of interstitial vortices between neighboring pinning rows.
Since the number of vortices in each interstitial row can vary over a
considerable range without destroying the row structure,
the maxima in $F^{Tr}_c$ are much broader 
than the peaks in $F^{L}_c$ associated with
longitudinal commensuration effects.
The row structures become increasingly anisotropic with increasing field until
a buckling transition occurs which marks the end of the maximum in $F^{Tr}_c$.
In Fig.~3(a) we illustrate the vortex positions
for $B/B_{\phi} = 8.0$  at a minimum in $F^{Tr}_c$
found at the point marked {\bf a} in Fig.~1(b).
The interstitial vortices form a mixture of two and three interstitial rows
between pinning site rows, producing the nonuniform trajectories shown
in Fig.~3(b).
At the maximum in $F^{Tr}_c$ marked {\bf c} in Fig.~1(b),
corresponding to 
$B/B_{\phi} = 10.67$, 
Fig.~3(c,d) shows 
that there are three well defined rows of
moving vortices between adjacent pinning site rows. 
Similarly, Fig.~3(e,f) indicates that there are four interstitial vortex rows
at $B/B_{\phi} = 17$, which falls on the maximum in $F^{Tr}_c$ at the point
marked {\bf e} in Fig.~1(b).
Near $B/B_{\phi} = 14$, where
$F^{Tr}_c$ passes through a minimum, 
the interstitial vortices form
a mixture of three and four rows, while for 
$B/B_{\phi} \gtrsim 19$ there is  
a mixture of four and five interstitial rows (not shown).  

Commensurability effects generated by the presence of an integer 
number of vortex rows 
between line-like barriers have been observed for
longitudinal vortex motion through channel geometries \cite{Kes,Anders} 
as well as critical currents in layered materials \cite{Griessen}, 
superconducting strips \cite{Souza}, and anisotropic pinning
arrays \cite{Karapetrov}. In all these cases the
commensurability occurs in the {\it static} vortex configurations.
This is distinct from the transverse depinning maxima that we observe
here, which arises due to commensurations in the
{\it dynamical} interstitial vortex configuration.  

\begin{figure}
\includegraphics[width=3.5in]{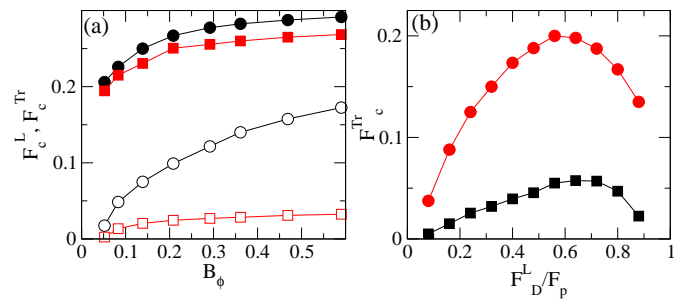}
\caption{ 
(a) The longitudinal critical depinning force $F^{L}_{c}$ vs $B_{\phi}$  
for $B/B_{\phi} = 2.0$ (filled squares)
and $B/B_{\phi} = 2.5$ (open squares)
and the transverse critical depinning force $F^{Tr}_c$ vs $B_\phi$
for $B/B_{\phi} = 2.0$ (filled circles)
and $B/B_{\phi} = 2.5$ (open circles)
(b) $F^{Tr}_{c}$ vs the applied
longitudinal force $F^{L}_{D}/F_{p}$ for $B/B_{\phi} = 3.0$ (filled circles) and 
$B/B_{\phi} = 3.67$ (filled squares).   
}
\end{figure}
      
As shown in Fig.~1(c), 
the transverse depinning threshold is much higher than the 
longitudinal depinning threshold. 
In Fig.~4(a) we quantify this effect by 
plotting $F_c^L$ and $F_c^{Tr}$ 
as a function of pinning density $B_\phi$
for a commensurate field $B/B_{\phi} = 2.0$ where 
$F_c^L$ passes through a peak and for an incommensurate
field $B/B_\phi=2.5$.
At the incommensurate field, $F_c^L$ 
and $F_c^{Tr}$ are both reduced.
At $B/B_{\phi} = 2.0$, $F_c^L$ increases monotonically with $B_\phi$
while $F^{Tr}_{c}$ shows a smaller increase; however, 
$F^{Tr}_c$ is significantly larger than $F^L_c$ over the entire range of
pinning densities studied.
At the incommensurate field $B/B_\phi=2.5$, we find 
a similar trend; however, $F^L_c$ 
increases much more slowly than $F^{Tr}_c$ with increasing $B_\phi$
and at $B_\phi=0.6$, $F^{Tr}_{c}$ is nearly an 
order of magnitude larger than $F^{L}_{c}$. 
In addition to increasing with increasing $B_\phi$,
the ratio $F^{Tr}_c/F^L_c$ increases with decreasing $B_\phi$ 
as $B_\phi$ approaches zero due to the
different rates at which the two thresholds approach zero.

The higher value of $F^{Tr}_{c}$ compared to $F^L_c$
can be understood by considering that 
longitudinal depinning occurs from the ground 
state configurations of the interstitial vortices 
\cite{Harada,Reichhardt} and    
is determined by 
the repulsive interactions between           
the vortices at the pinning sites and the interstitial vortices. 
In the ground state, the interstitial vortices occupy
positions that lower the repulsion from the pinned vortices, and 
the initial longitudinal depinning 
occurs when the interstitial vortices begin to move between the pinning 
sites, such as in Fig.~2(a). 
For the transverse depinning, when the
interstitial vortices are moving 
at a sufficiently high velocity in the longitudinal direction 
they do not have time to slip between the pinned vortices 
in the transverse direction, but instead come into close proximity with the
pinned vortices and interact strongly with them,
resulting in a high repulsive barrier for depinning.
If the longitudinal drive is set to 
a lower value before the transverse drive is applied, 
the interstitial vortices have more time to pass between 
the pinned vortices 
and $F^{Tr}_c$ decreases.
In Fig.~4(b) we illustrate this effect by
plotting $F^{Tr}_{c}$ versus $F^{L}_{D}/F_{p}$ 
for $B/B_{\phi} = 3.0$ and $B/B_{\phi} = 3.67$. 
In both cases 
$F^{Tr}_{c}$ increases from a low value with increasing $F^L_D/F_p$
until reaching a maximum value 
at $F^L_D/F_p=1.07$ for $B/B_\phi=3.0$ and at $F^L_D/F_p=1.12$
for $B/B_\phi=3.67$. 
Above this drive, $F^{Tr}_c$ decreases with increasing
$F^L_D/F_p$ as $F^L_D/F_p$ approaches 1
since the vortices at the pinning sites begin to depin for the higher
longitudinal drives, reducing the magnitude of the transverse critical
current.
      
In summary, we have shown that a new type of 
dynamical commensurability effect can occur 
for vortices in periodic pinning arrays. 
When interstitial vortices are moving between 
pinned vortices and an additional transverse force is applied, 
there is a finite transverse
critical depinning force which oscillates with field. 
The oscillation is not simply related to the 
matching of the vortices with the number of pinning sites 
as in the case for the longitudinal depinning,      
but is associated with the dynamical structure of the vortices which
allows for integer or non-integer numbers of rows of moving 
interstitial vortices between adjacent rows of pinning sites.
The transverse commensurability effects are much broader than those seen 
for the longitudinal depinning 
and each maximum in the transverse depinning force
can span several matching fields. Remarkably, the
transverse depinning force can be more than an order of magnitude 
larger than the longitudinal depinning force
due to the fact that the moving interstitial vortices 
are unable to move between the 
pinned vortices without coming close to the pinned vortices, 
which creates a strong repulsive barrier for transverse depinning.

This work was carried out under the auspices of the 
NNSA of the 
U.S. DoE
at 
LANL
under Contract No.
DE-AC52-06NA25396.

\end{document}